\documentclass[preprint,aps,showpacs,preprintnumbers,amsmath,amssymb,nofootinbib,graphicx]
{revtex4}
\usepackage{epsfig}
\begin{document}

\begin{flushright}
\end{flushright}

\newcommand{\be}{\begin{equation}}
\newcommand{\ee}{\end{equation}}
\newcommand{\bea}{\begin{eqnarray}}
\newcommand{\eea}{\end{eqnarray}}
\newcommand{\nn}{\nonumber}

\title{\large Possibility of observing Leptonic CP violation with perturbed Democratic mixing patterns}
\author{K. N. Deepthi, R. Mohanta }
\affiliation{
School of Physics, University of Hyderabad, Hyderabad - 500 046, India }

\begin{abstract}
The Daya Bay oscillation has recently reported the precise measurement of $\theta_{13}\simeq 8.8^\circ \pm 0.8^\circ $
or $\theta_{13} \neq 0$ at $5.2 \sigma$ level.
The observed non-zero $\theta_{13}$ can be accommodated by some general modifications to the Democratic mixing
matrix. Using such  matrices we study the possibility of observing non-zero CP violation in the
leptonic sector.
\end{abstract}

\maketitle

\section{Introduction}

The observation of neutrino oscillations has revealed that neutrinos
may have non-zero masses and lepton flavors are mixed.
Thus, analogous to the quark mixing, the  three
flavor eigenstates of neutrinos ($\nu_e,~ \nu_\mu,~ \nu_\tau$)
are related to the corresponding
mass eigenstates ($\nu_1,~ \nu_2,~ \nu_3$) by the unitary transformation
\bea
\left( \begin{array}{c}
 \nu_e       \\
\nu_\mu \\
\nu_\tau \\
\end{array}
\right ) \; =\; \left ( \begin{array}{ccc}
V_{e1}      & V_{e2}    & V_{e3} \\
V_{\mu 1}      & V_{\mu 2}    & V_{\mu 3} \\
V_{\tau 1}      & V_{\tau 2}    & V_{\tau 3} \\
\end{array}
\right ) \left ( \begin{array}{c}
\nu_1 \\
\nu_2 \\
\nu_3 \\
\end{array}
\right ),
\eea
where $V$ is the $3 \times 3 $ unitary
matrix known as PMNS matrix \cite{pmns,pmns1}, which contains three mixing angles
and three CP violating phases (one Dirac two Majorana phases).
The mixing matrix $V$ can be parameterized in terms of  three mixing angles $\theta_{12}^{}$,
$\theta_{23}^{}$, $\theta_{13}^{}$ and three CP-violating phases
$\delta, \rho, \sigma$  \cite{ref2} as
\begin{eqnarray}
V = \left( \begin{array}{ccc} c^{}_{12} c^{}_{13} & s^{}_{12}
c^{}_{13} & s^{}_{13} e^{-i\delta} \\ -s^{}_{12} c^{}_{23} -
c^{}_{12} s^{}_{13} s^{}_{23} e^{i\delta} & c^{}_{12} c^{}_{23} -
s^{}_{12} s^{}_{13} s^{}_{23} e^{i\delta} & c^{}_{13} s^{}_{23} \\
s^{}_{12} s^{}_{23} - c^{}_{12} s^{}_{13} c^{}_{23} e^{i\delta} &
-c^{}_{12} s^{}_{23} - s^{}_{12} s^{}_{13} c^{}_{23} e^{i\delta} &
c^{}_{13} c^{}_{23} \end{array} \right) P^{}_\nu \;
,\label{mat}
\end{eqnarray}
where $c^{}_{ij}\equiv \cos \theta^{}_{ij}$, $s^{}_{ij} \equiv \sin
\theta^{}_{ij}$ and $P_\nu^{} \equiv \{ e^{i\rho}, e^{i\sigma}, 1\}$ is
a diagonal matrix with CP violating Majorana phases $\rho$ and $\sigma$.
The global analysis of the recent results of various neutrino oscillation experiments
\cite{ref3} suggest the neutrino masses and mixing parameters at $1 \sigma$
level to be
\begin{eqnarray}
&&\Delta m^2_{21} = 7.59^{+0.20}_{-0.18})  \times 10^{-5}
\mbox{eV}^2\;,\nonumber\\
&&\Delta m_{31}^2 =
\left\{\begin{array}{ll}
+\left (2.50^{+0.09}_{-0.16}\right ) \times 10^{-3}_{}~~\mbox{eV}^2 &
~~{\rm for~normal~hierarchy~(NH)} \\
-\left (2.40_{-0.09}^{+0.08}\right ) \times 10^{-3}_{}~~\mbox{eV}^2 &
~~{\rm for~inverted~hierarchy~(IH)}
\end{array}\right. \nonumber \\
&&\sin^2 \theta_{12} = 0.312_{-0.015}^{+0.017}\;,\;\;\sin^2 \theta_{23}
= 0.52 \pm 0.06\;,\nn\\
&& \sin^2 \theta_{13}=0.013_{-0.005}^{+0.007}~~~~{\rm (NH)}\;\;~~
\sin^2 \theta_{13}=0.016_{-0.006}^{+0.008}~~~~{\rm (IH)}.
\end{eqnarray}
Furthermore, evidence for  $\theta_{13}\neq 0^\circ$ at about  $3\sigma$ level has been
obtained in a global analysis \cite{fogli}. In 2011, data from ${\rm T2K} $ \cite{t2k}, MINOS \cite{minos} and
Double Chooz experiment \cite{de} ruled out, for the first time,
$\theta_{13}=0$ at $3 \sigma$ level.

The Daya Bay Collaboration \cite{dayabay} has recently reported the first precise measurement of $\theta_{13}$ from
the reactor $\bar\nu_e \to \bar\nu_e$ oscillations. The best fit ($ 1 \sigma $) result is
\be
\sin^2 2 \theta_{13}=0.092 \pm 0.016({\rm stat}) \pm 0.005({\rm syst}),
\ee
which is equivalent to $\theta_{13} \simeq 8.8^\circ \pm 0.8^\circ $ or $ \theta_{13} \neq 0$ at $5.2 \sigma$ level.
This is followed by the results from  RENO collaboration \cite{reno}
\be
\sin^2 2 \theta_{13}=0.113 \pm 0.013({\rm stat}) \pm 0.019({\rm syst}).
\ee
These exciting observations imply that the smallest neutrino mixing angle is not really small
and the PMNS mixing matrix $V$ is not strongly hierarchical.
Evidence of non-zero reactor angle $\theta_{13}$  yields a potentially measurable $CP$ phase
$\delta $ in future neutrino oscillation experiments.
Thus, one of the important implications
of such observation  is that
 leptonic CP violation could be observable,
analogous to the observed CP violation in the quark sector.

The purpose of the present paper is to look for the possible existence of CP violation in the
lepton sector. The strength of CP violation in neutrino oscillations is described
by the Jarlskog rephasing  invariant \cite{jal}
\be
J={\rm Im}(V_{e1}V_{\mu 2}V_{e2}^* V_{\mu 1}^*) = {\rm Im}(V_{e2} V_{\mu 3}V_{e 3}^* V_{\mu 2}^*) = \cdots =
c_{12} s_{12} c_{13}^2 s_{13} c_{23} s_{23} \sin \delta\;,
\ee
which is proportional to the sine of the smallest mixing angle $\theta_{13}$.
In the quark sector the corresponding Jarlskog invariant is found to be $J_q \sim {\cal O}( 10^{-5})$
which is attributed to the strongly suppressed values of the quark flavor mixing angles.
In the lepton sector, since there are two large mixing angles, it could be possible to achieve
a relatively large $J$,  if
the CP violating phase is not vanishingly small.

\section{Methodology}

It has been shown in  Ref. \cite{rf10} that perturbations to various well-known mixing patterns,
i.e.,  bimaximal (BM) \cite{bi,bi1,bi2,bi3,bi4,bi5}, tri-bimaximal (TB)
\cite{tri,tri1,tri2,tri3,tri4,tri5,tri6,tri7,tri8,tri9} and
democratic mixing pattern (DC) \cite{minzhu,minzhu1},  would lead to neutrino mixing angles as given by current neutrino experiments.
However, in Ref. \cite{xing} it is very elaborately discussed that out of the five  typical mixing
patterns, i.e., the democratic, bimaximal,
tri-bimaximal, golden ratio and hexagonal forms,  the democratic mixing pattern provides a more natural
perturbation matrix, which can be obtained easily from either the flavour symmetry breaking or quantum corrections.
For a more clear illustration let us consider the mixing matrix to have the form
\be
V= (V_0 + \Delta V) P_\nu \ee
in which the leading term $V_0$ is a constant matrix responsible for two larger mixing angles $\theta_{12}$ and
$\theta_{23}$ and the correction  term $\Delta V$ is a perturbation matrix responsible for both the smallest mixing angle
$\theta_{13}$ and the Dirac CP violating phase $\delta$. Considering $V_0$ to be the Democratic mixing matrix $V_{\rm DC}$:
\begin{eqnarray}
V_{\rm DC} & = & \left ( \begin{array}{ccc}
\sqrt{\frac{1}{2}}&\sqrt{\frac{1}{2}}&0\\
\sqrt{\frac{1}{6}}&-\sqrt{\frac{1}{6}}&-\sqrt{\frac{2}{3}}\\
-\sqrt{\frac{1}{3}}&\sqrt{\frac{1}{3}}&-\sqrt{\frac{1}{3}}
\end{array}
\right )\;. \label{vtri}
\end{eqnarray}
the three mixing angles are found to be $\theta_{12}^{(0)}=45^\circ$, $\theta_{13}^{(0)}=0^\circ$ and
$\theta_{23}^{(0)} =\arctan (\sqrt 2) \simeq 54.7^\circ$.
Therefore, the corrections to all the large mixing angles $\theta_{12}$ and $\theta_{23}$ are
found to be
\be
\theta^* \equiv \theta_{12}^{(0)}-\theta_{12}=\theta_{23}^{(0)}-\theta_{23}
\approx 10^\circ.
\ee
This value of $\theta^*$ is
quite interesting as it is very close the observed non-zero $\theta_{13}$.

Motivated by the success of the DC mixing matrix in explaining the nonzero $\theta_{13}$,
we would like to scrutinize further the implication of this
mixing pattern. It is well known that the Democratic mass matrix is one of the most interesting candidate for the texture
of quark and charged-lepton mass matrices, since it naturally explains why the third generation particles
are much heavier than the first two generations. The democratic mixing matrix $V_{\rm DC}$ was originally obtained,
as the leading term of the lepton-flavor mixing matrix $V_{\rm PMNS}$, from the breaking of $S(3)_L \times S(3)_R$
flavor symmetry of the charged lepton mass matrix in the basis where the neutrino mass matrix is diagonal \cite{minzhu,minzhu1}.
First we will briefly review the modified DC mixing pattern and the resulting consequences.
The general modification of the mixing matrix \cite{rf10}, which will give nonzero $\theta_{13}$ could
be one of the following forms
\begin{eqnarray}
&&1.~ V_{\rm PMNS}= V_{\rm DC}^{} \cdot V_{ij}^{} \; , \nn\\
&&2.~  V_{\rm PMNS}= V_{ij}^{} \cdot  V_{\rm DC}^{} \; ,\nn\\
&& 3.~V_{\rm PMNS}= V_{\rm DC}^{} \cdot V_{ij}^{} \cdot V^{}_{kl} \; ,\nn\\
&& 4. ~ V_{\rm PMNS} =  V_{ij}^{} \cdot V^{}_{kl} \cdot
V_{\rm DC}^{} \; ,\label{p4}
\end{eqnarray}
where  $(ij), (kl) =(12), (13),
(23)$ respectively. The perturbation mixing matrices
$V^{}_{ij}$ are given by
\begin{eqnarray}
V^{}_{12} &=& \left (\begin{array}{ccc}
\cos x & \sin x &0\\
-\sin x &\cos x &0\\
0&0&1
\end{array}
\right )\;, \;\;~~~~~~~~~~~~~ V^{}_{23} = \left (\begin{array}{ccc}
1&0&0\\
0&\cos y  &\sin y ~e^{i \delta'}\\
0&-\sin y ~e^{-i \delta'}& \cos y
\end{array}\right )\nn\\
V^{}_{13} & = & \left ( \begin{array}{ccc}
\cos z &0&\sin z ~ e^{i\delta'} \\
0&1&0\\
-\sin z ~e^{-i \delta'} &0& \cos z
\end{array}
\right )\;. \label{vb}
\end{eqnarray}
In Ref. \cite{rf10}, it has been shown that out of the eighteen possible forms only five will accommodate the observed neutrino
oscillation data. These five forms are listed below:
\begin{eqnarray}
&& i.~~~ V_{\rm DC} V_{13} V_{12}\nn\\
&& ii.~~~ V_{\rm DC}V_{23}V_{13}\nn\\
&& iii.~~ V_{23}V_{13}V_{\rm DC}\nn\\
&& iv.~~~ V_{23}V_{12}V_{\rm DC}\nn\\
&& v.~~~ V_{13}V_{12}V_{\rm DC}
\end{eqnarray}
The implications of these five forms are extensively studied in Ref. \cite{rm},  if one considers texture one-zero mass matrices with vanishing CP violating phases.
In this paper we would like to investigate in detail these  mixing patterns
and their implications towards the observation of CP violation in the neutrino sector
without assuming the Dirac type CP violating phase to be zero.

\subsection{Case 1: $ V= V_{\rm DC} V_{13} V_{12} $}

Now we will consider the case where the PMNS matrix takes the form $ V= V_{\rm DC} V_{13} V_{12} $, which yields
\begin{eqnarray}
V = \displaystyle{\left( \begin{array}{ccc} \frac{c^{}_{x} c^{}_{z}-s^{}_{x}}{\sqrt{2}} & \frac{c^{}_{x}+c^{}_{z}s^{}_{x}}{\sqrt{2}} & \frac{s^{}_{z}}{\sqrt{2}} e^{i\delta^{'}} \\ \frac{s^{}_{x}}{\sqrt{6}}+c^{}_{x}(\frac{c^{}_{z}}{\sqrt{6}}+\sqrt{\frac{2}{3}}s^{}_{z}e^{i\delta^{'}})
 & \frac{-c^{}_{x}}{\sqrt{6}}+s^{}_{x}(\frac{c^{}_{z}}{\sqrt{6}}+\sqrt{\frac{2}{3}}s^{}_{z}e^{i\delta^{'}})
 & -\sqrt{\frac{2}{3}}c^{}_{z}+\frac{s^{}_{z}}{\sqrt{6}}e^{i\delta^{'}} \\
-\frac{s^{}_{x}}{\sqrt{3}}+c^{}_{x}(-\frac{c^{}_{z}}{\sqrt{3}}+\frac{s^{}_{z}}{\sqrt{3}}e^{i\delta^{'}}) &
\frac{c^{}_{x}}{\sqrt{3}}+s^{}_{x}(-\frac{c^{}_{z}}{\sqrt{3}}+\frac{s^{}_{z}}{\sqrt{3}}e^{i\delta^{'}}) &
-\frac{c^{}_{z}}{\sqrt{3}}-\frac{s^{}_{z}}{\sqrt{3}}e^{i\delta^{'}} \end{array} \right)},
\end{eqnarray}

By comparing the above matrix with the standard PMNS matrix (1) one can obtain the values
of the perturbation parameters $x$ and $z$. For illustration let us compare the (1,3)
element of both the matrices which gives us
 \begin{equation}
s_{13}e^{-i\delta}=\frac{s_{z}}{\sqrt{2}}e^{i\delta{'}}\;.
\end{equation}
Taking the modulus on both sides one can obtain the value of $z$. For numerical analysis we will use the
the values of  mixing angles from \cite{ref3,dayabay} as
 \be
 \sin^{2}\theta_{12}= 0.312^{+0.017}_{-0.015},~~~
~~\sin^{2}\theta_{23}= 0.52\pm 0.06, ~~~~~\sin^{2}\theta_{13}=0.023\pm 0.004.
\ee
 Using these input parameters, one can obtain obtain $z= (12.5\pm 1.1)^\circ$.
Similarly one can obtain the value of $x$ from the ratio of the elements $V(1,2)$ and $V(1,1)$ as
\begin{equation}
\tan{x}=\frac{c_z\tan{\theta_{12}}-1}{c_z+\tan{\theta_{12}}}\;,
\end{equation}
which gives us  $x=(-11.71\pm 0.12)^\circ$. To constrain the Dirac-type CP violating phase, we take the ratio of $V(2,3)$ and $V(3,3)$
elements, which gives
\begin{equation}
\cos \delta = \frac{c_z(\sqrt 2 - \tan \theta_{23})}{\sin \theta_{13} (1+ \sqrt 2 \tan \theta_{23})}
\end{equation}
Now varying the value of the mixing angle $\theta_{23}$  within  $1 \sigma$ range, we show in Figure-1 the correlation
plot between the Dirac type CP violating phase $\delta$ and $\theta_{13}$. From the figure it can be seen that the
allowed range of $\delta$ will be between $(0-60)^\circ$.

\begin{figure}[htb]
   \centerline{\epsfysize 2.5 truein \epsfbox{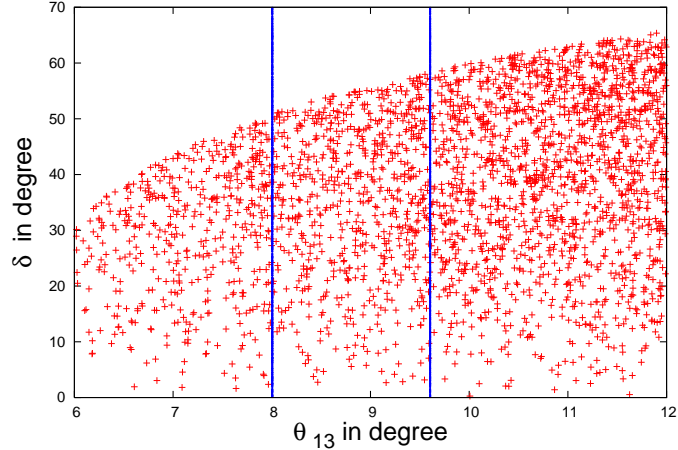}}
 \caption{
  The correlation plot between the mixing angle $\theta_{13}$ and Dirac  CP violating
  phase $\delta$. The vertical lines represent the $1-\sigma$ allowed range of
  $\theta_{13}$.}
  \end{figure}

The Jarlskog invariant is found to be
 \begin{equation}
J= \frac{1}{3\sqrt{2}}s_{z}(c_{2x}c_{z}-s_{2x}\frac{s^{2}_{z}}{2}) \sin{\delta'}=
\frac{1}{3}s_{13}(c_{2x}c_{z}-s_{2x}\frac{s^{2}_{z}}{2}) \sin{\delta}=(0.046 \pm 0.004) \sin{\delta}\;.
\end{equation}
Thus, substituting the value of $\delta$ as obtained from the correlation plot, we  expect
the CP violation parameter $J$ could be $J \leq 0.04$.

Now to evaluate the effective electron neutrino mass $m_{ee}$ that appears in neutrino less double beta decay,
we work in the basis where the charged leptons are diagonal and extract the (1,1) matrix element of
the rotated neutrino mass matrix given as \cite{rm}
\bea
m_{ee} = |m_{1}V_{11}^2+m_{2}V^{2}_{12}+m_{3}V^{2}_{13}|\;.
\eea
Assuming normal hierarchy structure of neutrino masses,
 we can eliminate the heavier neutrino masses $m_2$ and $m_3$ in terms of
the lightest neutrino mass $m_1$ and the observed mass square differences as
\bea
m_2 &=& \sqrt{m_1^2 + \Delta m_{21}^2}\nn\\
m_3 &=& \sqrt{m_1^2 + \Delta m_{31}^2 }\;,
\eea
we show in Figure-2, the variation of $m_{ee}$ with $m_1$, where the other parameters are
allowed to vary within their $1-\sigma$ range. Thus,for $m_1$ below ${\cal O}(10^{-2})$ eV, we get $m_{ee}  \leq 3.4 \times 10^{-2} $ eV.

\begin{figure}[htb]
   \centerline{\epsfysize 2.5 truein \epsfbox{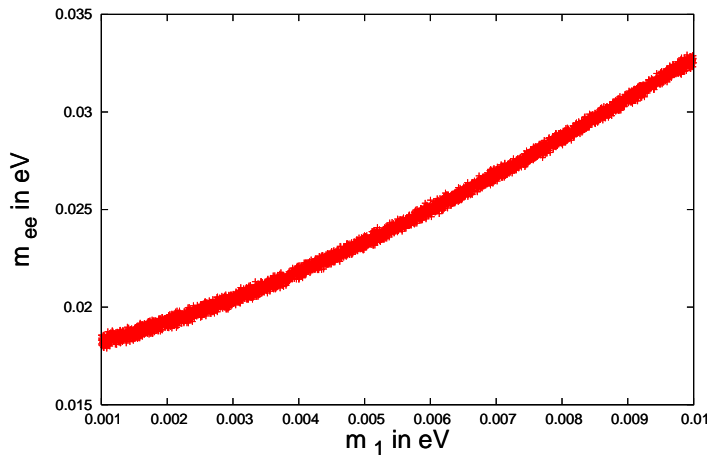}}
 \caption{
  Variation of $m_{ee}$ with the lightest neutrino mass $m_1$ for the NH scenario.}
  \end{figure}

\subsection{ Case 2: \textbf{$V= V_{\rm DC}V_{23}V_{13}$}}

Here we will consider the mixing matrix of the form $V= V_{\rm DC}V_{23}V_{13}$, which gives
\begin{eqnarray}
V= \left( \begin{array}{ccc} \frac{1}{\sqrt 2} \left (c_{z}-s_{y}s_{z}e^{2i\delta{'}}\right ) & \frac{c^{}_{y}}{\sqrt{2}} & \left (\frac{c^{}_{z}s^{}_{y}+s^{}_{z}}{\sqrt{2}}\right )e^{i\delta^{'}}\\ \frac{c^{}_{z}}{\sqrt{6}}+ s_z e^{i\delta^{'}}\left (\sqrt{\frac{2}{3}}c^{}_{y}+\frac{s^{}_{y}}{\sqrt{6}}e^{i\delta^{'}}\right ) & -\frac{c^{}_{y}}{\sqrt{6}}+\sqrt{\frac{2}{3}}s^{}_{y} e^{i\delta^{'}} & -c^{}_{z}\left (\sqrt{\frac{2}{3}}c^{}_{y}+\frac{s^{}_{y}}{\sqrt{6}}e^{i\delta^{'}} \right )+\frac{s^{}_{z}}{\sqrt{6}}e^{i\delta^{'}} \\ -\frac{c^{}_{z}}{\sqrt{3}}-s^{}_{z}(-\frac{c^{}_{y}}{\sqrt{3}}+\frac{s^{}_{y}}{\sqrt{3}}e^{i\delta^{'}})e^{i\delta^{'}} & \frac{1}{\sqrt 3} (c_{y}+s_{y}e^{i\delta^{'}}) & c^{}_{z}(-\frac{c^{}_{y}}{\sqrt{3}}+\frac{s^{}_{y}}{\sqrt{3}}e^{i\delta^{'}})-\frac{s^{}_{z}}{\sqrt{3}}e^{i\delta^{'}} \end{array} \right)\;.\label{eq24}
\end{eqnarray}
Comparing (1,2) element of above matrix with that of (2) we obtain
\begin{equation}
\cos{y}=\sqrt{2} \sin{\theta_{12}}\cos{\theta_{13}}\;.\label{eq25}
\end{equation}
 Substituting the values for $\theta_{12}$ and  $\theta_{13}$  in (\ref{eq25}) we obtain $ y=(38.66\pm 0.15)^{\circ}$.
 Similarly comparing the (1,3) element of (2) and (\ref{eq24}), and doing some
 algebraic manipulation  we obtain $z=-(21.5 \pm 0.8)^\circ$.

Proceeding in the same manner as in the previous case we obtain the CP violating phase
 $\delta$ and $s_{13}$ as
\begin{equation}
\cos{\delta}=\left(\frac{c_{z}s_{y}+s_{z}}{c_{z}s_{y}-s_{z}}\right)
\left(\frac{\tan(\theta_{23})-\sqrt{2}}{\sqrt{2}\tan(\theta_{23})+1}\right)\frac{c_{y}c_{z}}{s_{13}}\;.
\end{equation}
Varying the parameters within their $1-\sigma$ range, the correlation plot between $\delta$ and $s_{13}$ is
shown in Figure-3, which shows that the allowed range of $\delta$ in this case is between $(96-105)^{\circ}$.
\begin{figure}[htb]
   \centerline{\epsfysize 2.5 truein \epsfbox{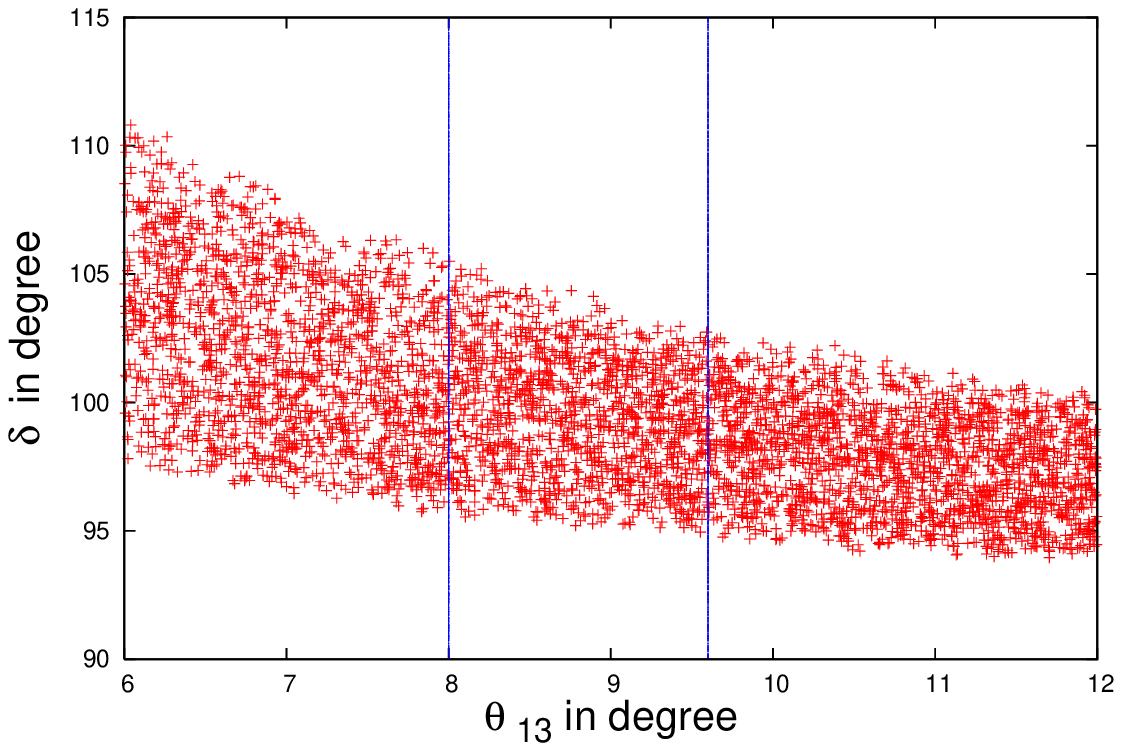}}
 \caption{
  The correlation plot between the mixing angle $\theta_{13}$ and Dirac  CP violating
  phase $\delta$. The vertical lines represent the $1-\sigma$ allowed range of
  $\theta_{13}$.}
  \end{figure}
The expression for CP-violation parameter $J$ is found to be
  \begin{equation} J = \sin{\delta{'}}\left(\frac{-s^{2}_{y}c_{y}s_{z}c_{z}}{6}+\frac{c^{2}_{y}s_{y}s^{2}_{z}}{6}
 -\frac{s^{3}_{y}c_{y}s^{2}_{z}}{6}\right)-\sin{2\delta{'}}\left(\frac{s^{2}_{y}c^{2}_{y}s^{2}_{z}}{3}\right) \;,
 \label{eq28}
\end{equation}
where $\delta'$ is related to $\delta $ through
\be
\cos \delta'=\left ( \frac{\sqrt 2 s_{13}}{c_z s_y+s_z} \right ) \cos \delta \;.
\ee
Now substituting $y$ and $z$ values in (\ref{eq28}) we obtain  $J\approx (0.024- 0.027) $.

\subsection{ Case 3: \textbf{$V=V_{23}V_{13}V_{\rm DC}$}}

In this case the mixing matrix $V$ is found to be
\begin{eqnarray}
V= \displaystyle{\left( \begin{array}{ccc} \frac{c^{}_{z}}{\sqrt{2}}-\frac{s^{}{z}}{\sqrt{3}}e^{i\delta{'}} & \frac{c^{}_{z}}{\sqrt{2}}+\frac{s^{}{z}}{\sqrt{3}}e^{i\delta^{'}} & -\frac{s^{}{z}}{\sqrt{3}}e^{i\delta{'}} \\ \frac{c_{y}-\sqrt{2}c_{z}s_{y}e^{i\delta{'}}-\sqrt{3}s_{y}s_{z}e^{2i\delta{'}}}{\sqrt{6}} & \frac{-c_{y}+\sqrt{2}c_{z}s_{y}e^{i\delta{'}}-\sqrt{3}s_{y}s_{z}e^{2i\delta{'}}}{\sqrt{6}} & \frac{-\sqrt{2}c_{y}-c_{z}s_{y}e^{i\delta{'}}}{\sqrt{3}} \\ -\frac{\sqrt{2}c_{y}c_{z}+s_{y}e^{i\delta{'}}+\sqrt{3}c_{y}s_{z}e^{i\delta{'}}}{\sqrt{6}} & \frac{\sqrt{2}c_{y}c_{z}+s_{y}e^{i\delta{'}}-\sqrt{3}c_{y}s_{z}e^{i\delta{'}}}{\sqrt{6}} & \frac{-c_{y}c_{z}+\sqrt{2}e^{i\delta{'}}s_{y}}{\sqrt{3}}\end{array} \right)}\;.\label{eq29}
\end{eqnarray}
As done in the previous cases from the confrontation of the elements of (2) with (\ref{eq29}) we obtain the perturbation angles
from
\begin{equation}
s_{13}e^{-i\delta}=\frac{s_{z}}{\sqrt{3}}e^{i\delta{'}}
\end{equation}
\begin{equation}
 \tan{y}=\frac{(c_{z}\tan{\theta_{23}}-\sqrt{2})}{(c_{z}-\sqrt{2}\tan{\theta_{23}})}
\end{equation}
$z= \left (15.22^{+1.32}_{-1.14}\right )^{\circ}$ and $y= (38.98\pm 0.10)^{\circ}$.

Expression for CP-violation phase is obtained by comparing the ratio of (2,3) and (3,3) elements of (2) and
 (\ref{eq29}) and replacing $\cos \delta'$ by
\begin{equation}
\cos{\delta{'}}=\frac{\sqrt{3}s_{13}\cos{\delta}}{s_{z}}\;,
\end{equation}
 as
\begin{equation}
\cos{\delta}=\frac{s_{z}c_{y}\tan{\theta_{23}}(c_{z}-\sqrt{2})}{\sqrt{3}s_{y}(\sqrt{2}\tan{\theta_{23}}
+c_{z})}\frac{1}{s_{13}}\;.
\end{equation}
In this case the allowed range of $\delta$ is found to be $(100-110)^{\circ}$.

The Jarlskog invariant parameter is found to be
\begin{eqnarray}
J &= & \sin{\delta{'}}\left(\frac{c^{2}_{y}s_{z}c_{z}}{3\sqrt{6}}-\frac{\sqrt{2}}{3}c_{z}s^{2}_{z}c_{y}s_{y}-
\frac{s^{2}_{z}c_{y}s_{y}c_{z}}{9\sqrt{2}}\right)\nn\\
&-& \sin{2\delta{'}}\left(\frac{c^{2}_{z}s_{z}c_{y}s_{y}}{3\sqrt{3}}-\frac{s^{2}_{y}s^{2}_{z}c^{2}_{z}}{6}\right)
+\sin{3\delta{'}}\left(\frac{s^{2}_{z}c_{z}s_{y}c_{y}}{3\sqrt{2}}\right)\;,
\end{eqnarray}
which yields  $J={\cal O}(10^{-3})$.

\subsection{ Case 4: \textbf{$ V= V_{23}V_{12}V_{DC}$}}

For $V= V_{23}V_{12}V_{DC}$ form, we get
\begin{eqnarray}
V=\displaystyle{\left( \begin{array}{ccc} \frac{c_{x}}{\sqrt{2}}+\frac{s_{x}}{\sqrt{6}} & \frac{c_{x}}{\sqrt{2}}-\frac{s_{x}}{\sqrt{6}} & -\sqrt{\frac{2}{3}}s_{x} \\ \frac{c_{x}c_{y}-\sqrt{3}c_{y}s_{x}-\sqrt{2}s_{y}e^{i\delta{'}}}{\sqrt{6}} & \frac{-c_{x}c_{y}-\sqrt{3}c_{y}s_{x}+\sqrt{2}s_{y}e^{i\delta{'}}}{\sqrt{6}} & -\frac{\sqrt{2}c_{x}c_{y}+s_{y}e^{i\delta{'}}}{\sqrt{3}} \\ \frac{-\sqrt{2}c_{y}-c_{x}s_{y}e^{i\delta{'}}+\sqrt{3}s_{x}s_{y}e^{i\delta{'}}}{\sqrt{6}} & \frac{\sqrt{2}c_{y}+c_{x}s_{y}e^{i\delta{'}}+\sqrt{3}s_{x}s_{y}e^{i\delta{'}}}{\sqrt{6}} & \frac{-c_{y}+\sqrt{2}c_{x}s_{y}e^{i\delta{'}}}{\sqrt{3}}
\end{array}\right)}.
\end{eqnarray}
In this case the perturbation angles are found to be $x=(10.7\pm 1.0)^\circ $ and $y=-\left (7.85^{+0.08}_{-0.13}\right )^\circ$ and one can obtain a correlation between $\theta_{13}$ and $\theta_{12}$ as
\begin{equation}
\sin \theta_{13}= \frac{1-\tan \theta_{12}}{\sqrt{2(1-\tan \theta_{12}+\tan^2 \theta_{12})}}\;.
\end{equation}
In this scenario, if we allow $\theta_{12}$ to vary in its $3 \sigma $ range a very narrow parameter space for
$\theta_{13}$ can be found in the $\theta_{12}-\theta_{13}$ plane. The Jarlskog Invariant is found to be
\begin{equation}
J=-\sqrt{\frac{2}{3}}s_{y}c_{y}s_{x}\left (\frac{c^{2}_{x}}{2}-\frac{s^{2}_{x}}{6}\right ) \sin{\delta{'}} = ((9.8 \pm 0.7)\times 10^{-3})\sin{\delta{'}}\;.
\end{equation}
However, it is not possible to constrain the CP violating phase $\delta$ in this case.

\subsection{ Case 5 : \textbf{$V=V_{13}V_{12}V_{DC}$}}

Here the mixing matrix is given as
\begin{eqnarray}
V=\displaystyle{\left( \begin{array}{ccc} \frac{\sqrt{3}c_{x}c_{z}+c_{z}s_{x}-\sqrt{2}s_{z}e^{i\delta{'}}}{\sqrt{6}} & \frac{\sqrt{3}c_{x}c_{z}-c_{z}s_{x}+\sqrt{2}s_{z}e^{i\delta{'}}}{\sqrt{6}} & -\frac{\sqrt{2}c_{z}s_{x}+s_{z}e^{i\delta{'}}}{\sqrt{3}} \\ \frac{c_{x}}{\sqrt{6}}-\frac{s_{x}}{\sqrt{2}} & -\frac{c_{x}}{\sqrt{6}}-\frac{s_{x}}{\sqrt{2}} & -\sqrt{\frac{2}{3}}c_{x} \\ -\frac{\sqrt{2}c_{z}+\sqrt{3}c_{x}s_{z}e^{i\delta{'}}+s_{x}s_{z}e^{i\delta{'}}}{\sqrt{6}} & \frac{\sqrt{2}c_{z}-\sqrt{3}c_{x}s_{z}e^{i\delta{'}}+s_{x}s_{z}e^{i\delta{'}}}{\sqrt{6}} & \frac{-c_{z}+\sqrt{2}s_{x}s_{z}e^{i\delta{'}}}{\sqrt{3}}
\end{array}\right)}\;.
\end{eqnarray}
The perturbation angles are $x=(151.3\pm 0.2)^\circ$ and $z=(28.8\pm 0.1)^\circ$ obtained from
\begin{equation}
-\sqrt{\frac{2}{3}}c_{x}=c_{13}s_{23}\;,
\end{equation}
and
\begin{equation}
 \tan{z}=\frac{\sqrt{3}c_{x}\tan{\theta_{12}}+s_{x}\tan{\theta_{12}}-\sqrt{3}c_{x}+s_{x}}{\sqrt{2}(1+\tan{\theta_{12}})}\;.
\end{equation}
As discussed in Ref. \cite{rf10}, this scenario also does not work for the best fit values and is valid only
 for a narrow region in the $\theta_{12}-\theta_{23}$ plane. The Jarlskog invariant is found to be
\begin{equation}
J=\left(\frac{c_{x}c_{z}s_{z}}{\sqrt{6}}-\frac{s_{z}c_{z}s_{x}}{\sqrt{2}}\right)\left(\frac{c^{2}_{x}}{3}-\frac{c_{x}s_{x}}{\sqrt{3}}\right)\sin{\delta{'}} = -(0.147\pm0.0007)\sin{\delta{'}}\;.
\end{equation}
Here also it is not possible to obtain the bounds on the CP violating phase $\delta$.

\section{Summary and Conclusion}
Among the three mixing angles of the neutrino mixing matrix, the
smallest reactor angle $\theta_{13}$ is the most important one to
understand the lepton mixing pattern completely. One of the main objective
of the currently running and upcoming neutrino experiments is to measure it
very precisely. The recent results from T2K, MINOS, Double Chooz, Daya Bay and RENO
experiments indicate non-zero and relatively largish $\theta_{13}$.
The fact that $\theta_{13}$ is not strongly suppressed  is certainly a good news to the experimental attempts towards
measurement of CP violation in the lepton sector.
Motivated by this  relatively large value of
the reactor mixing angle $\theta_{13}$,  we have studied the possibility of
observing leptonic CP violation for  a class of modified democratic mixing
patterns in a systematic way. It has been shown that a largish $\theta_{13}$
can be accommodated by a general modification of the democratic
mixing matrix.  The perturbations are of the form of Euler rotation angles and the perturbation
parameters can be determined by using the known values of the neutrino mixing angles.
If the neutrino mass matrix satisfies texture one-zero pattern, then out of the eighteen
such possible modifications only five are physically allowed. We consider those five
modified DC mixing matrices and study their implications on possible observation of CP
violation in the neutrino sector. We also obtain some non-trivial correlation
between $\theta_{13}$ and the Dirac CP violating phase $\delta$.
We have also estimated the leptonic CP violation  as well as the effective
electron neutrino mass $m_{ee}$ which  involves in neutrino less double beta decay
process. Our result indicates that it could be possible to observe
such CP violation effect in the upcoming long baseline experiments.

{\bf Acknowledgments}
KND  would like to thank University Grants Commission for financial support.
The work of RM was partly supported by the Council of Scientific and Industrial Research,
Government of India through grant No. 03(1190)/11/EMR-II.

\end{document}